\documentclass[10pt]{llncs}
\usepackage{amsmath,amssymb,amsfonts}
\usepackage{graphicx}
\usepackage{mathptmx} 
\let\llncssubparagraph\subparagraph
\let\subparagraph\paragraph
\usepackage[compact]{titlesec}
\let\subparagraph\llncssubparagraph
\usepackage{booktabs} 
\usepackage{array, makecell} 
\usepackage{makecell} 
\usepackage{multirow} 
\usepackage[font=small]{caption} %
\usepackage{textcomp} 
\usepackage{color} 
\newcommand{\quotes}[1]{``#1''} 

\usepackage{msc} 
\usepackage{url}
\usepackage{hyperref}
\usepackage{breakurl} 
\hypersetup{breaklinks=true}
\usepackage{adjustbox} 
\usepackage{wrapfig} 
\usepackage{sistyle} 
\SIthousandsep{,} 
\newcommand*\justify{%
	\fontdimen2\font=0.4em
	\fontdimen3\font=0.2em
	\fontdimen4\font=0.1em
	\fontdimen7\font=0.1em
	\hyphenchar\font=`\-
}
\usepackage{paralist} 
\usepackage{lipsum}  
\usepackage[misc]{ifsym} 
\usepackage{color, colortbl} 
\definecolor{lightYellow}{cmyk}{0,0.04,4.4,0}
\definecolor{lightPurple}{cmyk}{0,0.27,0,0.07}
\begin{document}

\author{Eman Salem Alashwali\inst{1,2} \and Pawel Szalachowski\inst{3} \and Andrew Martin\inst{1}} 
\institute{University of Oxford, Oxford, United Kingdom \\
	\email{\{eman.alashwali,andrew.martin\}@cs.ox.ac.uk} \\
	\and King Abdulaziz University (KAU), Jeddah, Saudi Arabia \\
	\and Singapore University of Technology and Design (SUTD), Singapore, Singapore\\
	\email{pawel@sutd.edu.sg}
}

\title{Towards Forward Secure Internet Traffic}
\maketitle
\setcounter{footnote}{0}

\begin{abstract} 
Forward Secrecy (FS) is a security property in key-exchange algorithms which guarantees that a compromise in the secrecy of a long-term private-key does not compromise the secrecy of past session keys. With a growing awareness of long-term mass surveillance programs by governments and others, FS has become widely regarded as a highly desirable property. This is particularly true in the TLS protocol, which is used to secure Internet communication. In this paper, we investigate FS in pre-TLS~1.3 protocols, which do not mandate FS, but still widely used today. We conduct an empirical analysis of over 10 million TLS servers from three different datasets using a novel heuristic approach. Using a modern TLS client handshake algorithms, our results show 5.37\% of top domains, 7.51\% of random domains, and 26.16\% of random IPs \textit{do not select} FS key-exchange algorithms. Surprisingly, 39.20\% of the top domains, 24.40\% of the random domains, and 14.46\% of the random IPs that \textit{do not select} FS, \textit{do support} FS. In light of this analysis, we discuss possible paths toward forward secure Internet traffic. As an improvement of the current state, we propose a new client-side mechanism that we call \quotes{Best Effort Forward Secrecy} (BEFS), and an extension of it that we call \quotes{Best Effort Forward Secrecy and Authenticated Encryption} (BESAFE), which aims to guide (force) misconfigured servers to FS using a best effort approach. Finally, within our analysis, we introduce a novel adversarial model that we call \quotes{discriminatory} adversary, which is applicable to the TLS protocol.
\end{abstract}

\section{Introduction}
\label{sec:intro}
\subsection{Problem}
Forward Secrecy (FS) is a security property in key-exchange algorithms which guarantees that a compromise in the secrecy of a long-term private-key does not compromise the secrecy of past session keys~\cite{menezes96}. With a growing awareness of long-term mass surveillance programs by governments and others, FS has become widely regarded as a highly desirable property. This is particularly true in the TLS protocol, which is used to secure Internet communication. Experience has shown the possibility of servers' long-term private-key compromise. For example, RSA~\cite{rivest78} long-term private-keys have been compromised through prime factorisation, due to advancement in computing power~\cite{cavallar00}\cite{kleinjung10}, or due to low entropy during keys generation~\cite{heninger12}. Furthermore, long-term private-keys can be compromised through implementation bugs as in the Heartbleed bug~\cite{hearbleed14}, through social engineering, or other attacks. Due to the increasing importance of FS, the new version of TLS, TLS~1.3, mandates it by design by prohibiting non-FS key-exchange algorithms~\cite{rescorla18tls13}. In recent years, it has been shown that some FS key-exchange algorithms (e.g. ECDHE) can achieve faster performance than non-FS (e.g. RSA) algorithms~\cite{huang14}. Despite recommendations to server administrators to select FS key-exchange algorithms, non-FS key-exchange algorithms are selected by more than 25\% of the servers in our IPs dataset as we will show later. As a result, clients proceed with non-FS key-exchange algorithms when connecting to these servers. This puts users' encrypted data at the risk of future decryption by adversaries who collect traffic today, and decrypt it whenever the targeted servers' private-key is compromised. Motivated by the importance of FS in Internet security, in this paper, we analyse the state of FS in pre-TLS~1.3 protocols, and discuss possible paths towards improving its adoption, including proposing a new best effort approach.

\subsection{Contribution}
Our contributions are as follows: first, we conduct an empirical analysis of FS on over 10 million TLS servers using a novel heuristic approach on three different datasets that contain top domains, random domains, and random IPs, which represent the real-world web. Unlike previous work that identifies servers that select non-FS key-exchange algorithms by capturing the servers' responses for TLS handshakes, our analysis employs a heuristic procedure that allows us to answer a deeper question: \textit{Do servers that select non-FS key-exchange algorithms support FS ones?} Our results provide new and useful insights to vendors, policy makers, and decision makers. Second, we discuss possible paths towards forward secure Internet traffic. Third, we propose a novel client-side mechanism that we call \quotes{Best Effort Forward Secrecy} (BEFS), and an extension of it that we call \quotes{Best Effort Forward Secrecy and Authenticated Encryption} (BESAFE), which aims to guide (force) misconfigured servers to FS key-exchange algorithms using a best effort approach. We implement and evaluate a proof-of-concept for it. Our mechanism adds value to the existing \quotes{all or nothing} approach. Finally, within our BEFS security analysis, we introduce a novel threat model that we call \quotes{discriminatory} adversary. The model is applicable to semi-trusted servers running protocols such as TLS that gives the server the power of selecting a security level, exemplified by the ciphersuite in our case, while the client has no means of verifying the server's actual capabilities (i.e. justifying the server's decision if it selects a non-preferred ciphersuite such as those that do not provide FS). We show how this power can be abused by semi-trusted servers to discriminate against their users for a powerful third-party's advantage (e.g. government intelligence), with minimal evidence and liabilities (e.g. legal) of the server's involvement in carrying out the attack.

\subsection{Scope}\label{sec:scope}
Our focus is pre-TLS~1.3 protocols, mainly the currently supported versions by most clients and servers, TLS~1.2, TLS~1.1, and TLS~1.0. As a shorthand, we refer to them as pre-TLS 1.3. TLS~1.2~\cite{rescorla08tls12} does not enforce FS by design, but still widely used today. As of April 2019, only 13.6\% of the top \num{150000} most popular domains support TLS~1.3, according to a report by SSL Labs~\cite{ssllabs19}. Furthermore, there are no known plans from standardisation bodies or browser vendors to deprecate TLS~1.2 yet. Although our work is mainly on pre-TLS~1.3, it has an impact on currently deployed systems.

\section{Background}
\label{sec:background}
\subsection{Transport Layer Security (TLS)}\label{sec:tls}
\begin{wrapfigure}{r}{0.5\textwidth}
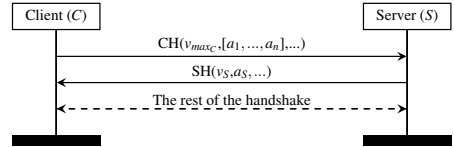

	\centering
	\vspace{-80pt}
	\resizebox{0.5\columnwidth}{!}{
		\setmsckeyword{} 
		\drawframe{no} 
		\begin{msc}[large values, /msc/level height=0.6cm, /msc/label distance=0.5ex , /msc/first level height=0.6cm, /msc/last level height=0.6cm, /msc/top head dist=0, /msc/bottom foot dist=0, /msc/environment distance=0, /msc/foot height=1.5ex]{}
			\setlength{\instwidth}{2\mscunit} 
			\setlength{\instdist}{6\mscunit} 
			\declinst{C}{}{Client ($C$)}
			\declinst{S}{}{Server ($S$)}
			
			\mess{CH($v_{max_C}$,[$a_1,...,a_n$],...)} {C}{S}
			\nextlevel
			
			\mess{SH($v_S$,$a_S,...$)} {S}{C}
			\nextlevel
			
			\mess*{The rest of the handshake}{C}{S}
			\mess*{}{S}{C}
			
		\end{msc}
	} 
	\vspace{-20pt}
	\caption{Illustration of the version and ciphersuite negotiation in pre-TLS~1.3 protocols.}
	\label{fig:TLS-CH} 	
	\vspace{-20pt}
\end{wrapfigure}
Transport Layer Security (TLS)~\cite{rescorla18tls13}\cite{rescorla08tls12} is one of the most important and widely used protocols to date. It is the main protocol used to secure Internet communication. TLS aims to provide data confidentiality, integrity, and authentication between two communicating parties. It has been in use since 1995, and was formerly known as the Secure Socket Layer (SSL). TLS consists of multiple sub-protocols including the TLS handshake protocol. In the handshake protocol, both communicating parties authenticate each other and negotiate security-sensitive parameters, including the protocol version and ciphersuite. The ciphersuite is an identifier that defines the cryptographic algorithms that, upon agreement between the communicating parties (client $C$ and server $S$), will be used to secure subsequent messages of the protocol. In pre-TLS~1.3, the ciphersuite defines the key-exchange, authentication, symmetric encryption, and hash algorithms. Some ciphersuites provide stronger security properties than others. For example, FS guarantees that a compromise in the server's long-term private-key does not compromise past session keys~\cite{menezes96}. Similarly, Authenticated Encryption (AE) provides confidentiality, integrity, and authenticity simultaneously, which provides stronger resilience against some attacks over the MAC-then-encrypt schemes~\cite{vaudenay02}\cite{alfardan13}. Most TLS clients today, such as mainstream web browsers, offer a mixture of ciphersuites that provides various levels of security such as FS, AE, both FS and AE, or none of them. The same applies to servers that select the session's ciphersuite.\par 

As depicted in \autoref{fig:TLS-CH}, at the beginning of a new TLS handshake, both communicating parties negotiate and agree on a protocol version and ciphersuite. The client sends a \texttt{ClientHello} (\texttt{CH}) message to the server. The \texttt{CH} contains several parameters including the client's maximum supported version $v_{max_C}$ and a list of ciphersuites [$a_1,...,a_n$] (we refer to them as the \textbf{client's offered versions} and the \textbf{client's offered ciphersuites}). Upon receiving the \texttt{CH}, the server selects a single version $v_S$ and a ciphersuite $a_S$ from the client's offer (we refer to them as the \textbf{server's selected version} and the \textbf{server's selected sciphersuite}), and responds with a \texttt{ServerHello} (\texttt{SH}) containing $v_S$ and $a_S$. If the server does not support the client's offered versions or ciphersuites, i.e. the client's offer is \textit{not} in the \textbf{server's supported versions} or the \textbf{server's supported ciphersuites}, the server responds with a handshake failure alert. 

\subsection{TLS Key-Exchange Algorithms}
There are two main key-exchange algorithms used in pre-TLS~1.3 protocols: the Rivest-Shamir-Adleman (RSA)~\cite{rivest78}, and the Ephemeral Diffie-Hellman (DHE)~\cite{whitfield76}. DHE has two variants: the Finite-Field (DHE) and the Elliptic-Curve (ECDHE). We use the term (EC)DHE to refer to either ECDHE or DHE. \par
\begin{figure}[!tp]
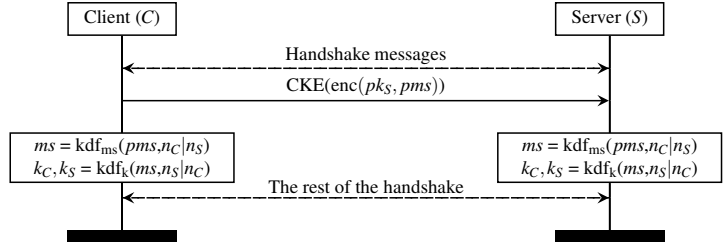
 
\centering
\resizebox{0.8\textwidth}{!}{
\setmsckeyword{} 
\drawframe{no} 
\begin{msc}[large values, /msc/level height=0.6cm, /msc/label distance=0.5ex , /msc/first level height=0.6cm, /msc/last level height=0.6cm, /msc/top head dist=0, /msc/bottom foot dist=0, /msc/environment distance=0, /msc/foot height=1.5ex]{}	
	\setlength{\instwidth}{2\mscunit} 
	\setlength{\instdist}{7\mscunit} 
			
	\declinst{C}{}{Client ($C$)}%
	\declinst{S}{}{Server ($S$)}%
			
	\mess*{Handshake messages}{C}{S}
	\mess*{}{S}{C}
			
	\nextlevel
	\mess{CKE(enc$(pk_S,pms)$)}{C}{S} 
			
	\nextlevel
	\action*{\parbox{4cm} {\centering $ms$ = kdf\textsubscript{ms}($pms$,$n_C|n_S$) \\ $k_{C},k_{S}$ = kdf\textsubscript{k}($ms$,$n_S|n_C$)}}{C}
			
	\action*{\parbox{4cm} {\centering $ms$ = kdf\textsubscript{ms}($pms$,$n_C|n_S$) \\ $k_{C},k_{S}$ = kdf\textsubscript{k}($ms$,$n_S|n_C$)}}{S}
	\nextlevel[2]
			
	\mess*{The rest of the handshake}{C}{S}
	\mess*{}{S}{C}
\end{msc}
} 
\vspace{-20pt}
\caption{Illustration of the RSA key-exchange in pre-TLS~1.3.}
\label{fig:TLS-RSA}
\vspace{-10pt}
\end{figure}

RSA Key-Exchange~\cite{rivest78} does not guarantee FS. As depicted in~\autoref{fig:TLS-RSA}, to generate a session key using RSA, the client generates a random value for the pre-master secret $pms$, encrypts it with the server's long-term RSA public-key $pk_S$ using the (enc) function, then sends it in a \texttt{ClientKeyExchange (CKE)} message. After that, both parties derive the master secret $ms$ from the $pms$ and their nonces $n_C$ and $n_S$, using a Key Derivation Function (kdf\textsubscript{ms}). Then, they compute the session keys $k_C$ and $k_S$ using the (kdf\textsubscript{k}) function. Clearly, the secrecy of the $pms$ relies on the secrecy of the server's long-term private-key $sk_S$ that is associated with the server's public-key $pk_S$ since every $pms$ is encrypted with the same server's long-term key $pk_S$ during the key's lifetime. Therefore, if the server's long-term private-key $sk_S$ is compromised at some point in the future, a passive adversary who has been collecting encrypted traffic, can recover the $pms$, and consequently, the $ms$, $k_C$ and $k_S$, and hence decrypt past sessions' encrypted data. \par 

\begin{figure}[!tp]
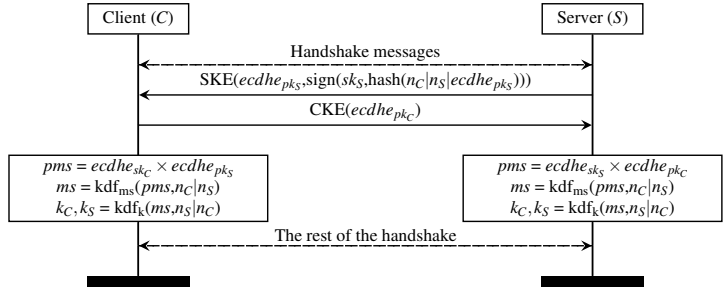
 
	\centering
	\resizebox{0.8\textwidth}{!}{
		\setmsckeyword{} 
		\drawframe{no} 
		\begin{msc}[large values, /msc/level height=0.6cm, /msc/label distance=0.5ex , /msc/first level height=0.6cm, /msc/last level height=0.6cm, /msc/top head dist=0, /msc/bottom foot dist=0, /msc/environment distance=0, /msc/foot height=1.5ex]{}
			
			\setlength{\instwidth}{2\mscunit} 
			\setlength{\instdist}{7\mscunit} 
			
			\declinst{C}{}{Client ($C$)}%
			\declinst{S}{}{Server ($S$)}%
			
			\mess*{Handshake messages}{C}{S}
			\mess*{}{S}{C}
			
			\nextlevel
			\mess{SKE($ecdhe_{pk_S}$,sign($sk_S$,hash($n_C|n_S|ecdhe_{pk_S}$)))} {S}{C}
			
			\nextlevel
			\mess{CKE($ecdhe_{pk_C}$)}{C}{S}
			
			\nextlevel		
			\action*{\parbox{5cm} {\centering $pms$ = $ecdhe_{sk_C} \times ecdhe_{pk_S}$ \\ $ms$ = kdf\textsubscript{ms}($pms$,$n_C|n_S$) \\ $k_{C},k_{S}$ = kdf\textsubscript{k}($ms$,$n_S|n_C$)}}{C}
			\action*{\parbox{5cm} {\centering  $pms$ = $ecdhe_{sk_S} \times ecdhe_{pk_C}$ \\ $ms$ = kdf\textsubscript{ms}($pms$,$n_C|n_S$) \\ $k_{C},k_{S}$ = kdf\textsubscript{k}($ms$,$n_S|n_C$)}}{S}
			\nextlevel[3]
			
			\mess*{The rest of the handshake}{C}{S}
			\mess*{}{S}{C}
		\end{msc}
	} 
	\vspace{-20pt}
	\caption{Illustration of the (EC)DHE key-exchange in pre-TLS~1.3.}
	\label{fig:TLS-ECDHE} 
	\vspace{-10pt}
\end{figure}
(EC)DHE~\cite{nir18}\cite{whitfield76} guarantees FS. As depicted in~\autoref{fig:TLS-ECDHE}, to generate a session key using (EC)DHE, the server sends its (EC)DHE public-key parameters $ecdhe_{pk_S}$, signed with its long-term private-key $sk_S$ using the (sign) function in a \texttt{ServerKeyExchange} (\texttt{SKE}) message. The client then sends its (EC)DHE public-key parameter $ecdhe_{pk_C}$ in a \texttt{ClientKeyExchange} (\texttt{CKE}) message. After that, both parties compute their $pms$, derive the $ms$ using the (kdf\textsubscript{ms}) function. Then, they compute the session keys $k_C$ and $k_S$ using the (kdf\textsubscript{k}) function. The (EC)DHE key is ephemeral, i.e. a fresh key is generated for each session. Unlike RSA, in (EC)DHE key-exchange, the $pms$ is not encrypted with the server's long-term key $pk_S$. Therefore, the $ms$, $k_C$ and $k_S$, do not rely on the secrecy of the server's long-term private-key $sk_S$. Hence, if the server's long-term private-key $sk_S$ is compromised at some point in the future, a passive adversary who has been collecting encrypted traffic, cannot recover the $ms$, $k_C$, and $k_S$ of past sessions.

\subsection{Terminology}\label{sec:terminology}
Throughout the paper, we use the term \textbf{FS-ciphersuites} to denote ciphersuites that support FS using the ECDHE key-exchange algorithm. We use the term \textbf{AE-ciphersuites} to denote ciphersuites that support AE using either the ChaCha20 stream cipher or the GCM mode of operation in the symmetric encryption algorithm. We use the term \textbf{FS+AE-ciphersuites} to denote ciphersuites that support both FS and AE using the ECDHE key-exchange algorithm and either the ChaCha20 stream cipher or the GCM mode of operation. Properties preceded with a \quotes{non} denotes a negated property. For example, the term \textbf{FS+non-AE-ciphersuites} denotes ciphersuites that support FS but not AE. These definitions are not meant for generalisation. They are limited to the paper's scope and to our experiment settings which are based on Google \texttt{Chrome}'s\footnote{As a shorthand, throughout the paper, we refer to Google \texttt{Chrome} as \texttt{Chrome}} ciphersuites. For example, \texttt{Chrome} only supports ECDHE to provide FS. Hence, our definition of FS-ciphersuites considers ECDHE only. To describe domains, we use the terms we defined in~\cite{alashwali19}: \textbf{main-domains} denotes domains consisting of a Top Level Domain (TLD) (e.g. \quotes{com}) prefixed by a single label, and do not have any further sub-domains, e.g. \quotes{example.com}; \textbf{plain-domains} denotes domains that are not prefixed with \quotes{www} sub-domains, e.g. \quotes{example.com}; and \textbf{www-domains} denotes domains that are prefixed with \quotes{www} sub-domains, e.g. \quotes{www.example.com}.

\section{Empirical Study}
\label{sec:empirical}
\subsection{Datasets}\label{sec:datasets}
We build three datasets that we name: \texttt{top-domains}, \texttt{random-domains}, and \texttt{random-ips}. We end up with \num{999884} distinct domains in the \texttt{top-domains} dataset, \num{4960390} distinct domains in the \texttt{random-domains} dataset, and \num{4881985} distinct IPv4 addresses in the \texttt{random-ips} dataset. The rationale behind choosing these three categories is to represent the real-world web as much as possible. In what follows, we explain how we build and pre-process each dataset.

\subsubsection{Top Domains Dataset} 
The \texttt{top-domains} dataset initial size is 1 million domains, obtained from the \texttt{Alexa} list of top 1 million most visited domains globally~\cite{alexa18}, retrieved on Aug. 22, 2018. We exclude the www-domains because we target plain-domains (see \autoref{sec:terminology} for our definitions of plain-domains and www-domains), which are the majority in the \texttt{Alexa} list. After excluding the www-domains, we end up with \num{999884} domains that are mainly (around 94.81\%) classified as main-domains.  

\subsubsection{Random Domains Dataset}
The \texttt{random-domains} dataset initial size is 5 million random domains obtained from a large dataset that contains \num{54063220} distinct (alphabetically unordered) domains that successfully completed a TLS handshake in Amann~et~al.~\cite{amann17}, which have been collected from multiple sources. To maintain consistency with the top domains dataset format, we extract 5 million domains from~\cite{amann17} that are classified as both plain-domains and main-domains. In this dataset, the TLDs scope is limited to generic TLDs (gTLDs), and does not include \quotes{multi-level} TLDs such as country-code TLDs (ccTLDs), e.g. \quotes{ac.uk}. This is to avoid the complexity of distinguishing domains that have sub-domains from domains that have ccTLDs, which is somewhat difficult to achieve with 100\% accuracy. To avoid repeated domains, from the 5 million random domains, we exclude the domains that exist in the top domains dataset, either \quotes{as is} or as a main-domain of a sub-domain in the top domains dataset (the top domains dataset contains a small percentage of sub-domains). We identify sub-domains in the top domain dataset in two steps: first, by using a regular expression, we extract the domains that have more than one dot \quotes{.}. Second, with the aid of \texttt{tldextract}~\cite{tldextract17} python library, we distinguish sub-domains from domains with country-code TLDs (ccTLD) such as \quotes{example.ac.uk} (the latter is considered a main-domain). We end up with \num{4960390} distinct random domains.      

\subsubsection{Random IPs Dataset}
The \texttt{random-ips} dataset initial size is 5 million distinct IPv4 addresses that have completed a successful TLS handshake with \texttt{Censys}, a search engine and database for servers and network devices on the Internet~\cite{censys15}, retrieved from the \texttt{Censys} IPv4 dataset on Oct. 20, 2018, through research access to the \texttt{Censys} database. To avoid repeated IPs, from the 5 million IPs, we exclude the IPs that are associated with any domain that has responded to a handshake in the scanning or inspection phases (further details on the scanning and inspection phases will be provided in the methodology in~\autoref{sec:methodology}). For this reason, we build the random IPs dataset after we finish the domains datasets scanning and inspection phases. We end up with \num{4881985} IPs. 

\subsection{Research Questions}\label{sec:questions}
Our analysis aims to answer the following main questions:
\begin{compactenum}
\item What is the percentage of servers that select non-FS-ciphersuites today?
\item Do servers that select non-FS-ciphersuites support FS-ciphersuites?
\item Do different dataset natures result in different trends in selecting and supporting FS-ciphersuites?
\end{compactenum}

Whilst addressing these main questions, the following side questions arose:
\begin{compactenum}
\item What is the percentage of servers that select FS+non-AE-ciphersuites after the client's FS-ciphersuites enforcement\footnote{The term \quotes{client's FS-ciphersuite enforcement} refers to a client that offers FS-ciphersuites exclusively. The same applies for \quotes{client's FS+AE-ciphersuite enforcement} but the latter offers FS+AE-ciphersuites exclusively. More details are provided next in the methodology in \autoref{sec:methodology}}? 
\item Do servers that select FS+non-AE-ciphersuites after the client's FS-ciphersuites enforcement support FS+AE-ciphersuites?
\item Can the client's FS-ciphersuites enforcement lead servers to lose the AE property?
\item Do servers that lose the AE property after the client's FS-ciphersuites enforcement support FS+AE-ciphersuites?
\end{compactenum}

\subsection{Methodology} \label{sec:methodology}
\begin{figure}[tp!]
\centering
\includegraphics[width=\textwidth]{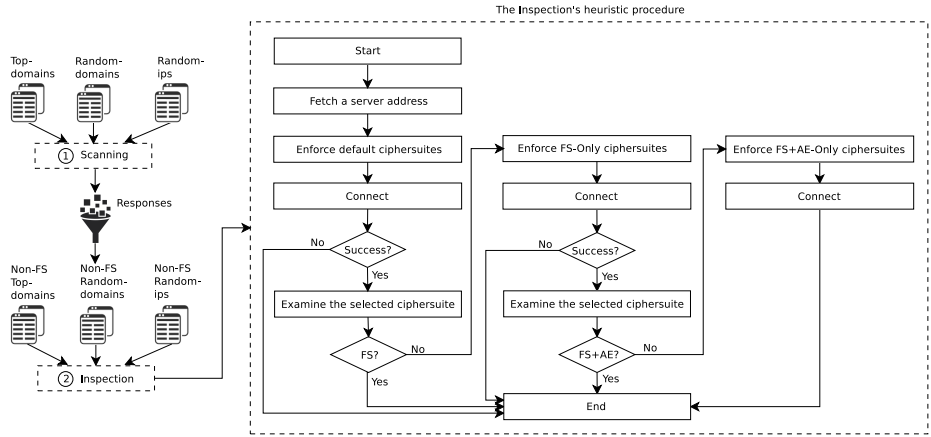}
\caption{A general overview of our methodology showing the two phases and their input.}
\label{fig:methodology}
\end{figure}

As depicted in~\autoref{fig:methodology}, our methodology consists of two main phases: a scanning phase followed by an inspection phase. 

\subsubsection{Scanning Phase} 
We consider the scanning phase as an exploration phase. In this phase, for each server address in our datasets, we perform a TLS handshake using the \texttt{tls-scan} tool~\cite{tls-scan16}, an open source fast TLS scanner capable of performing concurrent TLS connections. We customise the \texttt{tls-scan} to utilise the \texttt{OpenSSL 1.1.0g} library, and to support \texttt{Chrome}'s latest version pre-TLS~1.3 ciphersuites, which support various ciphersuites that provide FS, or AE, or none of them. We choose to base the scanning client on \texttt{Chrome}'s ciphersuites because \texttt{Chrome} is the most representative TLS client on the Internet. As of Feb. 2019, \texttt{Chrome}'s usage is 79.7\%~\cite{w3schools18}. \texttt{tls-scan} includes the Server Name Indication (SNI)\footnote{The SNI extension passes the domain name in the TLS handshake in order to obtain more accurate responses in virtual hosting environments, where a single server can host multiple domains~\cite{eastlake11}.} extension for domain name scans by default. We set the timeout argument to 5 seconds, and the concurrency argument to 50 connections. We ran the scans between Aug. 23, 2018 and Oct. 21, 2018 at the University of Oxford in discrete intervals based on the dataset. 

\subsubsection{Inspection Phase} \label{sec:inspection}
After the scanning phase is complete, we extract the responding addresses that selected non-FS-ciphersuites in the scanning phase. Each dataset is inspected within a maximum of 48 hours after its scanning phase is complete. For the inspection phase, we develop a TLS client that implements our heuristic procedure (see~\autoref{fig:methodology}), which works as follows, for each server's address:

\begin{compactenum}
\item The client performs a TLS handshake based on \texttt{Chrome}'s pre-TLS~1.3 default ciphersuites. This first handshake of the inspection phase is similar to the scanning phase handshake\footnote{Except that our inspection client does not support SSL~v3 while the scanning \texttt{tls-scan} client supports SSL~v3. However, we analyse FS regardless of the client's supported versions.}. The inspection's first handshake serves as a confirmation of the server's selected ciphersuite. It records the server's selected ciphersuite from a default client's view just before the heuristic procedure starts. If the handshake failed, the client records the error, and the heuristic procedure ends here.

\item Upon receiving the server's response to the first handshake (step 1), the client checks the server's selected ciphersuite: if it is a FS-ciphersuite, this means that the server has changed its behaviour after the scanning phase since all the inspection input addresses are for servers that selected non-FS-ciphersuites in the scanning phase. The client records the server's response, and the heuristic procedure for this server ends here. Otherwise, if the server's selected ciphersuite is still a non-FS-ciphersuite, we classify this server as \quotes{\textbf{stable}}, i.e. consistently selects a non-FS-ciphersuite. The client then updates its TLS context to support FS-ciphersuites \textit{exclusively}. The set of FS-ciphersuites may or may not support AE, i.e. it contains FS+AE-ciphersuites and FS+non-AE-ciphersuites. This context is more restricted than the default one. 

\item The client then performs a second handshake utilising the new FS-ciphersuites context. If the handshake failed, the client records the error and the heuristic procedure for this server ends here.

\item Upon receiving the server's response to the second handshake (step 3), the client checks the server's selected ciphersuite: if it is a FS+AE-ciphersuite, this means that the server supports FS+AE-ciphersuite after the client's FS-ciphersuites enforcement. The client then records the server's response, and the heuristic procedure for this server ends here. Otherwise, if the server's selected ciphersuite is a FS+non-AE-ciphersuite, the client updates its context to support FS+AE-ciphersuites \textit{exclusively}. This context is more restricted than the FS-ciphersuites context. 

\item The client then performs a third handshake utilising the new FS+AE-ciphersuites context. If the handshake failed, the client records the error and the heuristic procedure ends here.

\item  Upon receiving the server's response, the client records the response. The heuristic procedure for this server ends here.
\end{compactenum}
	
We develop and run the inspection client using \texttt{python 3.6.5}. Similar to the \texttt{tls-scan} client in the scanning phase, it utilises \texttt{OpenSSL 1.1.0g}. We enable the SNI for the top and random domains inspection (the IPs dataset do not need the SNI), and we set the timeout to 5 seconds. The results are then stored and analysed using MySQL database and queries.

\subsubsection{Identifying Device Types}
We classify device types into two categories: ordinary web servers and network devices. We use the term \quotes{\textbf{network device}} to refer to non-ordinary TLS servers, e.g. embedded web servers in network devices such as routers. To identify the device type, we input the IPs of the dataset in question in a query to \texttt{Censys} database to get the IPs metadata. We then produce a breakdown of the responding IPs grouped by the device type. We base our device types queries on the IPs, i.e. in the domains datasets, we first extract the distinct IPs behind the domains, because \texttt{Censys} is mainly an IP-based engine. We query the \texttt{Censys} snapshot that dates to the starting date of the scan or inspection (depending on the phase) of the dataset in question. \texttt{Censys} labels the device type of the network devices that it identifies, e.g. \quotes{DSL/cable modem}. If the device type field is empty, this means that the device is either an ordinary web server, or a network device that \texttt{Censys} cannot identify. Finally, we do not always obtain 100\% responses for the IPs that we query their metadata from \texttt{Censys}. However, overall, the percentages of the responses that we receive are between 98.36\% to 100\% (depending on the dataset) of the IPs we query.     

\subsubsection{Ethical Considerations} 
Our study is in line with the ethical recommendations in carrying out measurement studies~\cite{partridge16}. First, we do not collect private data. Second, we do not perform an exhaustive number of handshakes on any single server. Our clients' handshakes can by no means be classified as
a Denial of Service (DoS) attack. Third, we use a designated public IPv4 address per scanning device instead of Network Address Translation (NAT), to avoid potential disturbance to other users in our institution's network if a server has blocked our scanning or inspection device's IP. Fourth, we use informative DNS names that contain \quotes{TLS probing} to help server administrators identify our devices' activity in their logs. Finally, we inform the IT and security teams in our institution where the empirical study has been conducted so they expect a high volume of outgoing connections from our experiment devices, and to expect some incoming blacklisted certificates from random servers.   

\subsection{Results} \label{sec:results}
\subsubsection{Scanning Phase}
In this phase, we input the servers' addresses in our datasets. The results of the scanning phase are summarised in~\autoref{tab:scanning}.

\begin{table*}[!tp]
	\centering
	\caption{Summary of the scanning results. Every additional indentation means that the percentages are computed out of the previous level results. The \quotes{\% Network devices} are computed over the responding IPs to \texttt{Censys} metadata query (exact numbers are provided in text).}
	\label{tab:scanning}
	\begin{adjustbox}{max width=\textwidth}
		\begin{tabular}{lrrrrrrr}
			\toprule
			& \multicolumn{6}{c}{\thead{Datasets}} \\
			\cline{2-7}
			& \multicolumn{2}{r}{\texttt{top-domains}} 	& \multicolumn{2}{r}{\texttt{random-domains}} & \multicolumn{2}{r}{\texttt{random-ips}} \\
			\midrule
			Dataset size	
			& \multicolumn{2}{r}{\num{999884}}	  & \multicolumn{2}{r}{\num{4960390}} & \multicolumn{2}{r}{\num{4881985}}	 \\
			\midrule 
			\quad Responding servers		
			& \num{814333}&(81.44\%) & \num{3221249}&(64.94\%) & \num{4477279}&(91.71\%) \\
			\quad \quad Distinct IPs			
			& \num{468346}&(57.51\%) & \num{690912}&(21.45\%)  & \num{4477279}&(100\%) \\
			\quad \quad \quad \% Network devices		
			& \num{466}&(0.10\%)	 & \num{1208}&(0.18\%)	   & \num{518988}&(11.59\%) \\
			\midrule
			\rowcolor{lightYellow}[\dimexpr\tabcolsep+0.1pt\relax] 
			\quad \quad Select non-FS     
			& \num{43756}&(5.37\%)	 & \num{241994}&(7.51\%)   & \num{1171101}&(26.16\%) \\
			\bottomrule		
		\end{tabular}
	\end{adjustbox}
\end{table*}		

\paragraph{\textbf{Responding servers}}
As illustrated in~\autoref{tab:scanning}, the highest percentage of responses is in random IPs (91.71\%), followed by top domains (81.44\%), and finally random domains (64.94\%). The response rate is influenced by the dataset category. Both the IPs and top domains datasets are recent. That is, the addresses in the IPs dataset have recently completed a TLS handshake with the \texttt{Censys} engine~\cite{censys15}, and TLS adoption in top domains is high. The low response rate in random domains (64.94\%) is very likely due to the dataset age. It is obtained from a previous study that was published in 2017~\cite{amann17}. Hence, many domains could have gone down since then.\par 

In terms of device types, from the responding top domains, there are \num{468346} (57.51\%) distinct IPs behind all the top domains. We receive metadata responses for \num{464191} (99.11\%) of them from the \texttt{Censys} database. Of those, only \num{466} (0.10\%) IPs are labeled as network devices. From the responding random domains, there are \num{690912} (21.45\%) distinct IPs behind them. We receive metadata responses for \num{686085} (99.30\%) of them. Of those, there are \num{1208} (0.18\%) labeled as network devices. From the responding random IPs, we receive metadata responses for all of them (100\%). Of those, there are \num{518988} (11.59\%) IPs labeled as networked devices. Clearly, the percentage of network devices in the random IPs is higher than that in the top and random domains.

\paragraph{\textbf{Servers that select non-FS-Ciphersuites}}
From the responding servers, we find 5.37\% of the top domains, 7.51\% of the random domains, and 26.16\% of the random IPs, select non-FS-ciphersuites. The lowest percentage is in the top domains, the highest is in the random IPs, while in the random domains, it is slightly higher than that in the top domains. The fact that the random IPs dataset has the highest percentage of network devices can be correlated to the high percentage of servers that select non-FS-ciphersuites. We can confirm this in the inspection phase when we look closer at the device types of those servers that select non-FS-ciphersuites.

\subsubsection{Inspection Phase}
In this phase, we input the addresses of servers that select non-FS-ciphersuites in the scanning phase. \autoref{tab:inspection} summarises the inspection phase results.

\begin{table*}[!tp]
\centering
\caption{Summary of the inspection results. Every additional indentation means the percentages are computed out of the previous level results. The input of the inspection phase is the servers that selected non-FS-ciphersuites in the scanning phase. The \quotes{\% Network devices} are computed over the responding IPs to \texttt{Censys} metadata query (exact numbers are provided in text).}
\label{tab:inspection}
\begin{adjustbox}{max width=\textwidth}
\begin{tabular}{lrrrrrr}
\toprule
& \multicolumn{6}{c}{\thead{Datasets}} \\
\cline{2-7}
& \multicolumn{2}{r}{\texttt{\makecell*[r]{non-FS\\top-domains}}}
 & \multicolumn{2}{r}{\texttt{\makecell*[r]{non-FS\\random-domain}}} 	& \multicolumn{2}{r}{\texttt{\makecell*[r]{non-FS\\random-ips}}} \\
\midrule
Dataset size				
& \multicolumn{2}{r}{\num{43756}} & \multicolumn{2}{r}{\num{241994}}			& \multicolumn{2}{r}{\num{1171101}} \\
\midrule 
\quad Responding servers		
& \num{43374}&(99.13\%)	   & \num{240519}&(99.39\%)	& \num{1111802}&(94.94\%) \\
\midrule
\rowcolor{lightYellow}[\dimexpr\tabcolsep+0.1pt\relax] 
\quad \quad Select non-FS (stable)     
& \num{43158}&(99.50\%)    & \num{240274}&(99.90\%)	& \num{1111174}&(99.94\%)\\
\quad \quad \quad Distinct IPs      
& \num{33474}&(77.56\%)	   & \num{61522}&(25.60\%)	& \num{1111174}&(100\%) \\
\quad \quad \quad \quad \% Network devices 
& \num{76}&(0.23\%)	   	   & \num{361}&(0.59\%)		& \num{434076}&(39.06\%) \\
\midrule
\rowcolor{lightYellow}[\dimexpr\tabcolsep+0.1pt\relax] 
\quad \quad \quad Support FS     
& \num{16916}&(39.20\%)	   & \num{58636}&(24.40\%)	& \num{160706}&(14.46\%)\\
\quad \quad \quad \quad Distinct IPs      
& \num{12545}&(74.16\%)	   & \num{13839}&(23.60\%)	& \num{160706}&(100\%) \\
\quad \quad \quad \quad \quad \% Network devices 
& \num{12}&(0.10\%)	   	   & \num{27}&(0.20\%)		& \num{1503}&(0.94\%)	\\
\midrule 		
\rowcolor{lightPurple}[\dimexpr\tabcolsep+0.1pt\relax] 
\quad \quad \quad \quad Select FS+non-AE     
& \num{10091}&(59.65\%)	   & \num{38583}&(65.80\%)	& \num{93566}&(58.22\%) \\
\midrule 		
\rowcolor{lightPurple}[\dimexpr\tabcolsep+0.1pt\relax] 
\quad \quad \quad \quad \quad Support FS+AE	
& \num{1629}&(16.14\%)	   & \num{1289}&(3.34\%)	& \num{24128}&(25.79\%) \\
\midrule			
\quad \quad \quad \quad \quad Lose AE 
& \num{2686}&(26.62\%)	   & \num{1768}&(4.58\%) 	& \num{12769}&(13.65\%) \\
\midrule 
\quad \quad \quad \quad \quad \quad Support FS+AE
& \num{45}&(1.68\%)	   	   & \num{91}&(5.15\%) 		& \num{4668}&(36.56\%)	\\
\bottomrule
\end{tabular}
\end{adjustbox}
\end{table*}

\paragraph{\textbf{Responding servers}}
As  \autoref{tab:inspection} illustrates, over 99\% of top and random domains, and 94.94\% of IPs that select non-FS-ciphersuites in the scanning phase, have responded to our inspection client's handshake. The low response rate in the IPs dataset compared to the top and random domains datasets is very likely attributed to SSL~v3 devices as our inspection client does not support SSL~v3, while the scanning client does. It is also very likely that those non-responding IPs are mostly for network devices since using legacy versions in network devices is more common than that in ordinary web servers~\cite{samarasinghe17}.   

\paragraph{\textbf{Servers that still select non-FS-Ciphersuites (stable)}}
In our work, we use the term \quotes{\textbf{stable}} to refer to servers that consistently select non-FS-ciphersuites in both the inspection's first handshake and the scanning handshake. As shown in~\autoref{tab:inspection}, clearly, the stability in selecting non-FS-ciphersuites among all datasets is high (over 99\% in all datasets), despite the difference in the supported protocol versions in the scanning and inspection clients (the scanning client supports SSLv3 while the inspection does not). This suggests that, to some extent, servers' selected ciphersuites are not affected by the negotiated versions. \par 

In terms of device types, out of the top domains that select non-FS-ciphersuites, there are \num{33474} (77.56\%) distinct IPs behind all these domains. Of those, we receive metadata responses for \num{33079} (98.82\%) IPs from the \texttt{Censys} database. Of those, there are \num{76} (0.23\%) IPs labeled as networked devices. Of the random domains that select non-FS-ciphersuites, there are \num{61522} (25.60\%) distinct IPs behind them. We receive metadata for \num{61176} (99.44\%) IPs from \texttt{Censys}. Of those, there are \num{361} (0.59\%) IPs labeled as networked devices. Of the random IPs that select non-FS, we receive metadata for \num{1111174} (100\%). Of those, there are \num{434076} (39.06\%) IPs labeled as networked devices. Network devices represent no more than 0.59\% of top and random domains that select non-FS-ciphersuites. However, more than a third of servers that select non-FS-ciphersuites in the random IPs dataset are labeled as network devices. The high percentage of network devices in the random IPs is likely the reason for the high percentage of servers that select non-FS-ciphersuites.    

\paragraph{\textbf{Servers that select non-FS-Ciphersuites, but support FS-Ciphersuites}}
We find 39.20\% of top domains, 24.40\% of random domains, 14.46\% of random IPs, that select non-FS-ciphersuites in the inspection phase, do support FS-ciphersuites. The top-domains are the highest, followed by the random domains, and finally, the random IPs are the lowest. Interestingly, this is a shifted paradigm for the percentages of servers that select non-FS-ciphersuites that is shown in~\autoref{tab:scanning}, where the random IPs have the highest percentage and the top domains have the lowest percentage. The results reflect that the lack of FS-ciphersuite selection in the top and random domains is to a large extent due to misconfiguration, while in the random IPs, it is mostly due to lack of support. \par 

In terms of device types, out of the top domains that select non-FS-ciphersuites but support FS-ciphersuites, there are \num{12545} (74.16\%) distinct IPs behind them. We receive metadata responses for \num{12339} (98.36\%) IPs. We find \num{12} (0.10\%) of them are labeled as networked devices. Of the random domains that select non-FS-ciphersuites but support FS-ciphersuites, there are \num{13839} (23.60\%) distinct IPs behind them. We receive metadata responses for \num{13744} (99.31\%) IPs. Of those, \num{27} (0.20\%) are labeled as network devices. Of the random IPs, that select non-FS-ciphersuites but support FS-ciphersuites, we receive metadata responses for \num{160706} (100\%) IPs. Of those, \num{1503} (0.94\%) are labeled as network devices. The results show that the majority of those devices are not identified as network devices, even in the random IPs dataset that shows the highest percentage of network devices. Those servers that select non-FS-ciphersuites and turned to support FS-ciphersuites are not network devices. Therefore, most of the network devices that select non-FS-ciphersuites, do not support FS-ciphersuites.  

\paragraph{\textbf{Servers that select FS+non-AE-Ciphersuites after enforcing FS-Ciphersuites}}
Out of the top domains, random domains, and random IPs that support FS-ciphersuites after enforcement (row label \quotes{Support FS} in \autoref{tab:inspection}), there are 59.65\% top domains, 65.80\% random domains, and 58.22\% random IPs that select FS+non-AE-ciphersuites. The reason for selecting non-AE can be attributed to the fact that TLS~1.0 and TLS~1.1 do not support AE-ciphersuites~\cite{salowey08}, and these devices might be running legacy versions of TLS. Otherwise, this is attributed to misconfiguration. To better understand this situation, we next check whether those servers support FS+AE-ciphersuites or not.

\paragraph{\textbf{Servers that select FS+non-AE-Ciphersuite after enforcing FS-Ciphersuites, but support FS+AE-ciphersuites}}
Of the top domains, random domains, and random IPs that select FS+non-AE-ciphersuites after FS-ciphersuites enforcement (row label \quotes{Select FS+non-AE} in \autoref{tab:inspection}), there are 16.14\% top domains, 3.34\% random domains, and 25.79\% random IPs, that support FS+AE-ciphersuite. At this point of the heuristic procedure, the majority of the IPs do not belong to network devices. The majority of the top and random domains that select FS+non-AE-ciphersuites do not support FS+AE-ciphersuites. However, selecting FS+non-AE-ciphersuites while supporting FS+AE-ciphersuites in the IPs dataset is the highest, which we classify as misconfiguration. 
 
\paragraph{\textbf{When enforcing FS-Ciphersuite causes losing the AE property}}
Of the top domains, random domains, and randm IPs that select FS+non-AE-ciphersuites after enforcing FS-ciphersuites (row label \quotes{Select FS+non-AE} in \autoref{tab:inspection}), there are 26.62\% top domains, 4.58\% random domains, and 13.65\% random IPs, were selecting AE before enforcing FS-ciphersuites, i.e. were selecting non-FS+AE-ciphersuites. This can be either because they do not support any FS+AE-ciphersuites, or due to misconfiguration. This can be clarified next.

\paragraph{\textbf{\justify Servers that lose the AE property after enforcing FS-Ciphersuites, but support FS+AE-ciphersuites}}
Out of the top domains, random domains, and random IPs that lose the AE property after enforcing FS (row label \quotes{Lose AE} in \autoref{tab:inspection}), we find 1.68\% top domains, 5.15\% random domains, 36.56\% random IPs do support FS+AE-ciphersuites. The results reflect that losing the AE property after enforcing the FS in the top and random domains is to a large extent due to a lack of support for FS+AE-ciphersuites, but in the random IPs, it is mostly due to misconfiguration.

\section {Towards Forward Secure Internet Traffic}
\label{sec:befs}
In this section, we discuss possible paths towards forward secure Internet traffic from a client's perspective. Then, we propose and evaluate a novel client-side mechanism that we call Best Effort Forward Secrecy (BEFS), and an extension of it that we call Best Effort Forward Secrecy and Authenticated Encryption (BESAFE). We choose to focus our discussion and solutions on clients because unlike servers, clients are controlled by few players, e.g. browser vendors. Client-side security enhancement mechanisms are easier to adopt, as shown by recent adoptions of client-side mechanisms such as Google's Certificate Transparency (CT)~\cite{ct19}, and others. 

\subsection{Deprecating non-FS-Ciphersuites in TLS Clients}
The most straight-forward approach towards forward secure Internet traffic is deprecating non-FS-ciphersuites from TLS clients. As a result, these clients will not be able to establish TLS connections with servers that do not support FS-ciphersuites. This is a conservative approach that has been taken by browser vendors and standardisation bodies in the past with some protocol versions and algorithms such as SSL~v3~\cite{barnes15_ssl3} and RC4~\cite{popov15}, after their insecurity has become clear. However, deprecating non-FS-ciphersuites \textit{now} can be more problematic than the case of deprecating SSL~v3 and RC4 in 2014 and 2016 respectively. By way of comparison, Lee et al.\footnote{Despite the study's age (conducted in 2006), to the best of our knowledge,~\cite{lee07} is the only study that tried to assess servers' supported ciphersuites prior to deprecating RC4 and SSL~v3. Note that identifying the \textit{supported} ciphersuites for a server is different from identifying the \textit{selected} ciphersuite. The former requires multiple handshakes, while the latter requires a single handshake, for each server.} conducted a survey in 2006 to asses the cryptographic strength of TLS servers~\cite{lee07}. It shows that 98.36\% of the surveyed servers support TLS~1.0, the latest version at the time of the study, and 57.17\%  of the servers support AES encryption, which was in its early years as it was standardised in 2001~\cite{aes01}. In light of these figures, we speculate that SSL~v3 and AES adoption when they were deprecated by most browsers in 2014 and 2016 respectively was over 99\%. On the other hand, in our results, we calculate an approximation of the servers that support FS-ciphersuites in each dataset. To this end, we first calculate the number of servers that select non-FS-ciphersuites and do not support FS-ciphersuite which can be derived from~\autoref{tab:inspection} by calculating (\quotes{Select non-FS (stable)}$-$\quotes{Support FS}), and then subtracting those results from the overall responses in \autoref{tab:scanning}'s row label \quotes{Responding servers}, which gives: \num{788091} (96.78\%) top domains, \num{3039611} (94.36\%) random domains, and \num{3526811} (78.77\%) random IPs. Our results are in line with \texttt{Censys} Oct. 26, 2018 snapshot figures that show 97.44\% of \texttt{Alexa}'s top domains, and 77.94\% of IPs in all IPv4 space, support FS-ciphersuite. However, our results are more accurate as we include not only servers that \textit{select} FS, but also servers that \textit{support} FS but select non-FS, which can be guided through client's enforcement as we will explain next. In addition, we utilises modern client ciphersuites. On the other hand, \texttt{Censys} only measures servers that \textit{select} FS, and utilises somewhat legacy ciphersuites. We conclude that the percentages of servers that support FS-ciphersuites are less than that in RC4 and SSL~v3 cases, especially in the IPs datasets. The lack of supporting FS-ciphersuites by those servers can be explained by the fact that until recent years, (EC)DHE key-exchange algorithms have been viewed as resource-exhaustive compared to RSA key-exchange, despite the fact that this argument is no longer true with the ECDHE variant as shown in~\cite{huang14}.      
 
\subsection{Best Effort Forward Secrecy (BEFS)}   
\subsubsection{Overview}\label{sec:overview}
The gist of our BEFS mechanism is guiding (forcing) misconfigured servers towards FS-ciphersuites. As explained in~\autoref{sec:tls}, in ordinary TLS clients such as web browsers, the client offers default ciphersuites, which includes FS-ciphersuites and non-FS-ciphersuites. Upon receiving the client's offer, a server that does not support or does not prefer to select a FS-ciphersuite will select a non-FS-ciphersuite, and sends its selected ciphersuite to the client. The client accepts the server's choice, and the rest of the communication proceeds with a non-FS-ciphersuite. On the other hand, in BEFS, we exploit the TLS ciphersuite negotiation dynamics to influence (bias) the server's choice towards FS-ciphersuites. That is, a BEFS-enabled client first offers FS-ciphersuites \textit{exclusively} [$a_{1_{fs}},...,a_{n_{fs}}$]. Upon receiving the client's offered ciphersuites, a server that supports FS-ciphersuites will be guided (forced) to select a FS-ciphersuite $a_{S_{fs}}$, even if it prefers to select a non-FS-ciphersuite, since FS-ciphersuites are the only offered ciphersuites as illustrated in~\autoref{fig:default}. As shown in~\autoref{tab:inspection}, of the servers that select non-FS-ciphersuites, there is between 14.46\% to 39.20\% that \textit{do support} FS-ciphersuites, which can benefit from the BEFS enforcement mechanism. If the server indeed does not support FS-ciphersuites, it will return a failure alert (see~\autoref{sec:tls} for a background on TLS version and ciphersuite negotiation). In this case, the BEFS-enabled client makes a second handshake utilising default ciphersuites [$a_{1_{fs}},...,a_{n_{nonfs}}$], which includes non-FS-ciphersuites in addition to the previously offered FS-ciphersuites as~\autoref{fig:befs} illustrates. Hence, a server that does not support FS-ciphersuites can still select a non-FS-ciphersuite $a_{S_{nonfs}}$ after the client falls back. BEFS can be viewed as a form of the \quotes{Opportunistic Security} concept~\cite{dukhovni14}, but at the FS property level. That is, it guides servers to select FS whenever they support it. 
   

\begin{figure}[!tp] 
\captionsetup{width=.4\textwidth, font=small}
\centering
\begin{minipage}{.45\textwidth}
\centering
\resizebox{\columnwidth}{!}{
\setmsckeyword{} 
\drawframe{no} 
\begin{msc}[large values, /msc/level height=0.6cm, /msc/label distance=0.5ex , /msc/first level height=0.6cm, /msc/last level height=0.6cm, /msc/top head dist=0, /msc/bottom foot dist=0, /msc/environment distance=0, /msc/foot height=1.5ex]{}
	\setlength{\instwidth}{2\mscunit} 
	\setlength{\instdist}{6\mscunit} 
				
	\declinst{C}{}{Client ($C$)}
	\declinst{S}{}{Server ($S$)}
				
	\mess{CH(...,[$a_{1_{fs}},...,a_{n_{fs}}$],...)} {C}{S}
	\nextlevel
				
	\msccomment[side=right, msccomment distance=0.75cm, /msc/every msccomment/.append style={text width=2.5cm}]{Prefers non-Fs; Supports FS}{S}
	\nextlevel 
				
	\mess{SH(...,$a_{S_{fs}}$,...)} {S}{C}
	\nextlevel
				
	\mess*{FS connection}{C}{S}
	\mess*{}{S}{C}
				
	\end{msc}
	} 
	\vspace{-10pt}	
	\caption{A BEFS-enabled client handshake when the server prefers to select a non-FS-ciphersuite while supporting FS-ciphersuites. The server is forced to select FS-ciphersuite through client FS-ciphersuite enforcement.}
	\label{fig:default} 
	\vspace{-10pt}
\end{minipage}
\begin{minipage}{.45\textwidth}
\centering
\resizebox{\columnwidth}{!}{
\setmsckeyword{} 
\drawframe{no} 
\begin{msc}[large values, /msc/level height=0.6cm, /msc/label distance=0.5ex , /msc/first level height=0.6cm, /msc/last level height=0.6cm, /msc/top head dist=0, /msc/bottom foot dist=0, /msc/environment distance=0, /msc/foot height=1.5ex]{}
	\setlength{\instwidth}{2\mscunit} 
	\setlength{\instdist}{6\mscunit} 
				
	\declinst{C}{}{Client ($C$)}
	\declinst{S}{}{Server ($S$)}
				
	\mess{CH(...,[$a_{1_{fs}},...,a_{n_{fs}}$],...)} {C}{S}
	\nextlevel
				
	\msccomment[side=right, msccomment distance=0.75cm, /msc/every msccomment/.append style={text width=3cm}]{Prefers non-Fs; Supports non-FS only}{S}
	\nextlevel 
				
	\mess{Error} {S}{C}
	\nextlevel
				
	\mess{CH(...,[$a_{1_{fs}},...,a_{n_{nonfs}}$],...)} {C}{S}
	\nextlevel
				
	\mess{SH(...,$a_{S_{nonfs}}$,...)} {S}{C}
	\nextlevel
				
	\mess*{non-FS connection}{C}{S}
	\mess*{}{S}{C}	
\end{msc}
} 
\vspace{-10pt}
\caption{A BEFS-enabled client handshake when the server does not support FS-ciphersuites. The client falls back to non-FS-ciphersuites only when the server indeed does not support FS.}
\label{fig:befs}
\vspace{-10pt} 
\end{minipage}
\end{figure}


\subsubsection{The Fallback} \label{sec:fallback}
We now address the fallback aspect. We define three categories of client-side fallbacks: silent fallback, interactive fallback, and signaled fallback. In what follows, we explain them in light of the BEFS mechanism.
\par{\textbf{Silent fallback.}} 
Silent fallbacks do not involve the user or the server. If used in BEFS, if the FS-ciphersuites handshake failed, the client falls back to default ciphersuites (which include non-FS-ciphersuites), in the background, and performs a second handshake utilising default ciphersuites. Silent fallbacks remove the security decision-making overhead from the user at the cost of security. Silent fallbacks do not provide security against active adversaries who can perform downgrade attacks (for a background on downgrade attacks, see~\cite{alashwali18}). BEFS with silent fallback is secure against passive adversaries, which adds a significant value in the case of FS. It makes mass surveillance more difficult to achieve as the adversary has to actively perform downgrade attacks for each session.

\par{\textbf{Interactive fallback.}}
Interactive fallbacks involve the user. If used in BEFS, when the FS-ciphersuites handshake fails, the client (e.g. web browser) presents an interrupting warning message and asks the user whether to proceed or not. If the user chooses to proceed, the client falls back from FS-ciphersuites to default ciphersuites and performs a second handshake. Otherwise, if the user chooses not to proceed, the client does not fall back, and aborts the TLS handshake. Interactive fallbacks provide security against active adversaries. Interactive fallbacks are similar to the widely-known self-signed certificate active warnings~\cite{akhawe13}. Active security warnings have been shown to be more effective than passive ones such as passive indicators that do not interrupt the user's task~\cite{akhawe13}. However, active security warnings have to be used with caution in order to not cause the habitation effect, where users ignore them because they see them too often~\cite{sunshine09}. Therefore, if the majority of servers that do not support FS-ciphersuites (i.e. those that require fallback) are network devices, interactive fallback can be acceptable, as these devices are normally visited infrequently by a limited number of users, such as the device's owner. 

\par{\textbf{Signaled fallback.}} 
Signaled fallbacks involve the server. Therefore, if they are not incorporated in the protocol by design, they require modifications or updates to the server, e.g. a patch to the TLS implementation, to enable the server from interpreting the client's signal. In signaled fallbacks, the client sends a signal, i.e. a special value, to inform the server that the client has performed a fallback. The server aborts the handshake if it is not expecting a fallback, e.g. in BEFS case, if the server supports FS-ciphersuites. Signaled fallbacks provide security against active adversaries, if we assume authenticated messages. Signaled fallbacks have been proposed in the TLS fallback Signaling Cipher Suite Value (SCSV)~\cite{moller14_scsv}. It has been used to mitigate TLS version downgrade attacks, mainly the POODLE attack~\cite{moller14}, and has been widely adopted as shown in~\cite{amann17}. In BEFS, our problem deals with misconfigured servers and less security-aware server administrators. Had they been security-aware, they would have configured their servers to select FS-ciphersuites. Therefore, in BEFS case, we do not consider signaled fallbacks as a solution that can be adopted quickly. Therefore, we do not include it in our analysis in the coming section.

\subsubsection{BEFS Security Analysis}
We now analyse the security of BEFS against three adversarial models: passive network adversary, active network adversary, and our newly introduced discriminatory adversary. 

\par{\textbf{Passive Network Adversary}}
Passive adversaries can collect network traffic, but cannot interfere (e.g. modify, inject, replay, or drop) protocol messages. They may obtain access to the server's long-term private-key at some point in the future. Once the server's long-term private-key is compromised, a passive adversary who has been collecting non-FS network traffic can now decrypt it. BEFS aims to ensure the selection of FS-ciphersuites whenever the server supports FS-ciphersuites. In FS-ciphersuites, an ephemeral key is generated for each session, and this key is not encrypted with the server's long-term private-key. By selecting FS-ciphersuites, if the server's long-term private-key is compromised, the adversary cannot compromise past session keys. In TLS, the (EC)DHE key-exchange algorithms are provably secure against passive adversaries~\cite{jager12}. Therefore, BEFS with all types of fallback mechanisms is secure against passive adversaries.
  
\par{\textbf{Active Network Adversary}}
Unlike passive adversaries, active adversaries can interfere with protocol messages, e.g. by modifying, injecting, replaying, or dropping  messages. Similar to the passive adversaries, they may obtain access to the server's long-term private-key at some point in the future, hence be able to decrypt non-FS-ciphersuite traffic. BEFS security against active adversaries can be analysed with the two fallback mechanisms explained earlier in~\autoref{sec:fallback}. First, in terms of BEFS with silent fallback, since the user of a BEFS-enabled client with silent fallback is not aware of the fallback, an active adversary can perform a downgrade attack by dropping the initial FS-ciphersuites handshake message to lead the client to fall back and perform a default handshake. Hence, misconfigured servers that select non-FS-ciphersuites but support FS-ciphersuite will not be guided, i.e. will select non-FS-ciphersuite, while with BEFS, they will be guided (forced) to select FS-ciphersuites instead. BEFS with silent fallback does not provide security against active adversaries. Second, we analyse BEFS with interactive fallback against an active adversary. This moves the security decision to the user. Users can be classified into two categories: security-aware users, who read the warning message and reject the fallback when they care about FS. The second category of users is less security-aware users, who will not do so. BEFS with interactive fallback and security-aware users is secure against active adversaries. The warning message content and the users' reactions to it are beyond the scope of this paper. BEFS with interactive fallbacks can find its application in special browser modes for sensitive communications, in the same vein of \texttt{Chrome}'s incognito and \texttt{Firefox} private modes, which are available for privacy-aware users.   
  
\par{\textbf{Discriminatory Adversary}}
The discriminatory adversary is located at the server and discriminates against its clients in terms of the security level it provides to them (FS-ciphersuite vs. non-FS-ciphersuite in our case). The discriminatory adversarial model is applicable to semi-trusted servers running protocols such as TLS, which gives the server the power of selecting some parameters that define the security level of a particular session, exemplified by the ciphersuite in our case, while the client has no means of verifying the server's actual capabilities, i.e. justifying the server's decision if it selects a non-preferred ciphersuite such as a non-FS-ciphersuite. This power can be abused by semi-trusted servers to discriminate against their users, for a powerful third-party's advantage. The discriminatory adversary can be compelled by, or collude with the third-party, such as government intelligence, to weaken the security of \textit{some} connections, e.g. those coming from specific geographic locations. In our case, the discriminatory adversary denies the FS property to some users, whilst enabling it for others. The discriminatory adversary (server) can then provides its long-term private-key that is used for digital signatures and non-FS session keys ($pms$) encryption to the powerful third-party, after the key's expiration, when it is no longer used by the server. This allows the third-party to decrypt the data of those users who have been discriminated against, but not the data of other users who have been provided with strong security, i.e. FS-ciphersuite in our case. This adversarial model gives the semi-trusted server several advantages compared to  giving every session key or the decrypted data itself to the third-party, which is impractical for servers to carry out, especially in the case of large-scale surveillance. Another advantage to the semi-trusted server lies in the minimal liabilities (e.g. legal) in being directly involved in leaking their users' data, or in giving their private-key to the adversary during the key's lifetime. Such an adversarial model is not far from the export-grade cryptography law that was mandated until the late 90s, where software vendors, for example, were compelled to weaken the security of software exported to outside the United States (US), to enable US intelligence from breaking their security. Furthermore, leaked confidential documents by Edward Snowden suggests similar scenarios, where giant companies collude with government intelligence by introducing backdoors that are known to, and can be exploited by, those powerful adversaries (e.g. the \quotes{PRISM} program)~\cite{wikipedia_prism_18}. \par 

Our discriminatory adversarial model is inspired by the \quotes{malicious-but-cautious}~\cite{ryan14} and the Secretly Embedded Trapdoor with Universal Protection (SETUP)~\cite{young96} adversarial models. The \quotes{malicious-but-cautious} model assumes a cloud service provider (server) can act maliciously but is cautious not to leave a verifiable trace of its malicious behaviour. However, it does not assume that the malicious server is willing to enable a third-party to access some users' data. On the other hand, the SETUP model assumes a cryptographic system (server in our case) can enable a third-party to secretly obtain secret information such as the private-key that decrypts the encrypted data from the system's encrypted output. Our discriminatory adversary weakens the security against some users for a third party's advantage, and is also cautious not to leave a verifiable trace of its malicious behavior, e.g. by selecting a supported but non-preferred ciphersuite (non-FS-ciphersuite), as it is still accepted by most clients for backward compatibility. \par 

To better analyse BEFS against the discriminatory adversary (server), we further classify this adversary into two variants: weak discriminatory and strong discriminatory. In the weak variant, the adversary submits to the client's offer (ciphersuites in our case). That is, if the client offers strong choices exclusively, the weak discriminatory has no choice but to select from them, mainly to avoid detection. In the strong variant, the adversary refuses to select strong choices, which forces the client to fallback in order to connect to the server. BEFS with silent fallback is secure against the weak discriminatory adversaries. However, strong discriminatory adversaries require interactive fallback and security-aware users. In today's real-world settings, the weak variant is more realistic. However, the strong variant can be detected through BEFS and security-aware users. Note that this analysis of BEFS against a discriminatory adversary is independent of considerations about the communication channel. That is, if an active adversary is present in the communication channel, interactive fallback is required with both variants of the discriminatory adversary, in order for BEFS to meet its security goal.  

\subsubsection{Best Effort Forward Secrecy and Authenticated Encryption (BESAFE)} \label{sec:besafe}
Given the fact that more than 50\% of the servers select FS+non-AE-ciphersuite after enforcing FS-ciphersuites and that between 16.14\% to 25.79\% of them support FS+AE-ciphersuites, as an extension to BEFS, we propose BESAFE which adds an additional step to enforce not only FS-ciphersuites, but also FS+AE-ciphersuites. This improvement adds an additional restriction: the client offers FS+AE-ciphersuites \textit{exclusively} at the first handshake attempt. If it failed, the client falls back to BEFS: it tries FS-ciphersuites \textit{exclusively}, and if it failed, it falls back to default ciphersuites. The BESAFE mechanism guides servers towards FS+AE-ciphersuites. Similar to BEFS, BESAFE is secure against passive adversaries, or weak discriminatory adversaries with all types of fallbacks, and against active adversaries, or strong discriminatory adversaries with interactive fallback and security-aware users.  

\subsubsection{BEFS and BESAFE Performance}\label{sec:evaluation}
\begin{wraptable}{r}{0.5\textwidth}
	\centering
	\vspace{-25pt}
	\caption{The BEFS and BESAFE mechanisms latency in ms compared to the default one when connecting to servers that do not support FS-ciphersuites.}
	\label{tab:performance}
	\centering
	\begin{adjustbox}{max width=\columnwidth}
		\begin{tabular}{lrrr} 
			\toprule
			TLS Client 	& \thead{Max.}  & \thead{Min.}  		& \thead{Avg.} \\
			\midrule
			\texttt{Default} 			
			&  4.10						& 0.64					& 1.69  		\\
			\texttt{BEFS-Enabled} 		
			&  5.34						& 1.79	 				& 3.47	  		\\
			\texttt{BESAFE-Enable}		
			&  8.60						& 3.27					& 5.19			\\
			\bottomrule
		\end{tabular}
	\end{adjustbox}
\vspace{-10pt}
\end{wraptable}
We measure the latency that BEFS and its extension BESAFE incur into a TLS connection establishment with domains that do not support FS-ciphersuites, i.e. when more than one attempt is performed to complete a TLS handshake (otherwise, in BEFS, if the server supports FS-ciphersuites, and in BESAFE, if the server supports FS+AE-ciphersuites, there will be a single handshake as normal and no additional latency is incurred). To this end, we extract 5000 top domains that do not support FS-ciphersuites from our results. We implement a TLS client that supports \texttt{Chrome}'s pre-TLS~1.3 default ciphersuites using \texttt{Python 3.6.5} and utilising \texttt{OpenSSL 1.1.0g}. We disable TLS certificate validation and session tickets (resumption), and enable the SNI. Our client performs three consecutively handshakes for each domain: Default, BEFS-enabled, and BESAFE-enabled handshakes. We run the client on a machine equipped with a 3.2 GHz \texttt{Intel Core i5} processor, 8 GB of RAM, and a 1000 Mbps wired Ethernet card that has a public IPv4 address at the University of Oxford. We measure the time to complete a TLS handshake in a socket connection in milliseconds using the \texttt{process\_time()}, a process-wide timer in python's \texttt{time} module. We count the domains that triggered BEFS and BESAFE to resort to default ciphersuites (i.e. do not support FS-ciphersuites), and also responded to the default TLS handshake. Then we extract the maximum, minimum and average time they take for each of the three handshake types. There are 4501 domains that do not support FS-ciphersuites and responded to the three types of handshakes we examine. The results based on these responses are summarised in~\autoref{tab:performance}. We can also infer the latency that BESAFE incurs into a connection to a server that does not support FS+AE-ciphersuites but supports FS+non-AE-ciphersuites from the BEFS latency (since both require two attempts).




\subsubsection{Improved Performance Through Parallel Attempts}
As shown in the previous section, BEFS introduces a latency on the default TLS connection establishment, but only if the server does not support FS-ciphersuites. To minimise this latency, the client can implement BEFS attempts in parallel instead of consecutive. That is, the client sends two \texttt{CH}s one with default ciphersuites and the second with FS-ciphersuites, in parallel to the server. For each TLS session establishment, the client waits for all the \texttt{CH} attempts' responses to return. If there is a valid response to the FS-ciphersuites attempt, the client proceeds with the FS-ciphersuites response. Otherwise, if the FS-ciphersuites attempt has failed, the client proceeds with the default ciphersuite response. The same applies to BESAFE but the client sends three handshakes and first checks the FS+AE-ciphersuites, then the FS-ciphersuites attempts' response, before deciding to proceed with default ciphersuites.

\section{Related work}
\label{sec:related}
Lee~et~al.~\cite{lee07} scanned around \num{19000} TLS servers based on top domains lists. They evaluate the cryptographic strengths in TLS servers including the supported key-exchange algorithms. In~\cite{holz11}, Holz~et~al. provide statistics for the most popular selected ciphersuite by TLS servers.  
More recently, Kotzias~et~al.~\cite{kotzias18} examined the impact of high-profile attacks on TLS deployment, and Calzavara~et~al.~\cite{calzavara19} analysed the \texttt{Alexa} top \num{10000} websites against known HTTPS vulnerabilities. All the aforementioned studies do not analyse the selected versus supported key-exchange algorithms as we do. In~\cite{huang14}, Huan~et~ al. provide an experimental study on TLS FS deployment on \num{473802} of \texttt{Alexa}'s top domains using an enumeration-based method. Our study analyses larger and more diverse datasets than that in~\cite{huang14}, using a novel heuristic approach. Additionally, our study provides a recent view on FS adoption over the one in~\cite{huang14}, which dates back to 2014. Apart from measurement studies, several studies show that RSA long-term private-keys can be compromised either due to advances in computing power, deployment, or implementation flaws. Kleinjung~et~al.~\cite{kleinjung10} and Cavallar~et~al.~\cite{cavallar00} show that 786-bit and 512-bit RSA keys can be factored using powerful machines. Heninger~et~al.~\cite{heninger12} conducted a measurement study, which is replicated by Alashwali~\cite{alashwali13}, that shows that factorable RSA keys are widespread in network devices on the Internet, due to low entropy during prime generation. The Heartlbleed bug~\cite{hearbleed14} in the \texttt{OpenSSL} TLS library shows that implementation bugs can cause private-key compromise.

\section{Conclusions}
\label{sec:conclusion}
In this paper, we analysed the state of FS on over 10 million servers on the Internet. Using a modern TLS client handshake algorithms, our results show 5.37\% of top domains, 7.51\% of random domains, and 26.16\% of random IPs \textit{do not select} FS key-exchange algorithms. Surprisingly, we found that 39.20\% of the top domains, 24.40\% of the random domains, and 14.46\% of the random IPs that \textit{do not select} FS, \textit{do support} FS. We then discussed possible paths towards FS. We showed that a best effort approach can add a value over the \quotes{all or nothing} approach and can increase FS or FS and AE adoption in misconfigured servers.

\section*{Acknowledgment}
We thank the \texttt{Censys} team~\cite{censys15}, the CS's IT and OxCERT teams at the University of Oxford, and the \texttt{tls-scan} developer, Binu Ramakrishnan, for technical support. Pawel's work was supported by the SUTD SRG ISTD 2017 128 grant.

\Urlmuskip=0mu plus 1mu\relax 
\bibliographystyle{splncs04}
\bibliography{ref}

\end{document}